\documentclass[]{aa}  

\def\rmit#1{{\it #1}}              %% italics (RR mode, Kluwer)
\def\specchar#1{{\sc #1}}

\def\Ha{\mbox{H\,$\alpha$}}
\def\Hb{\mbox{H\,$\beta$}}
\def\CaIIH{\mbox{Ca\,\specchar{ii}\,\,H}}       %% use \CaIIK\ for space
\def\CaIIK{\mbox{Ca\,\specchar{ii}\,\,K}} 
\def\CaII{\mbox{Ca\,\specchar{ii}}} 

\def\arcsec{\hbox{$^{\prime\prime}$}}

\def\eg{\rmit{e.g.}}

\def\arcsec{\hbox{$^{\prime\prime}$}}

\usepackage{graphicx}
\usepackage{xcolor}
\usepackage{txfonts}
\usepackage[colorlinks=false]{hyperref}
\usepackage{soul} %allows the use of strikethrough command: \st{}

\begin{document}

%   \title{Fluctuations in Magnetic Field in \ion{He}{I} 10830 \AA{} in sunspot umbra and umbral dots}

\title{Observations of umbral flashes in the resonant sunspot chromosphere}

   \author{T. Felipe,
          \inst{1,2}\fnmsep\thanks{tobias@iac.es}
          S. J. Gonz\'alez Manrique,
          \inst{1,2,3}
          D. Mart\'inez-G\'omez,
          \inst{1,2}
          M. M. G\'omez-M\'iguez,
          \inst{1,2}
          E. Khomenko,
          \inst{1,2}
          C. Quintero Noda,
          \inst{1,2}
           \and
          H. Socas-Navarro\inst{1,2}
          }

   \institute{Instituto de Astrof\'{\i}sica de Canarias 
              38205 C/ V\'{\i}a L{\'a}ctea, s/n, La Laguna, Tenerife, Spain
         \and
             Departamento de Astrof\'{\i}sica, Universidad de La Laguna
             38205, La Laguna, Tenerife, Spain 
         \and
             Leibniz-Institut für Sonnenphysik (KIS), Schöneckstr. 6, 79104 Freiburg, Germany
%         \and  
%             Astronomical Institute, Slovak Academy of Sciences, 05960 Tatranská Lomnica, Slovak Republic\\
             }

   \date{Received ; accepted }

 \titlerunning{Umbral flashes in a chromospheric resonant cavity}
  \authorrunning{Felipe et al.}
  
% \abstract{}{}{}{}{} 
% 5 {} token are mandatory
 
  \abstract
   {In sunspot umbrae, the core of some chromospheric lines exhibits periodic brightness enhancements known as umbral flashes. The consensus is that they are produced by the upward propagation of shock waves. This view has recently been challenged by the detection of downflowing umbral flashes and the confirmation of the existence of a resonant cavity above sunspots.}
   {We aim to determine waves' propagating or standing nature in the low umbral chromosphere and confirm or refute the existence of downflowing umbral flashes. }
   {Spectroscopic temporal series of \CaII\ 8542 \AA, \CaIIH, and \Ha\ in a sunspot were acquired with the Swedish Solar Telescope. The \Ha\ velocity was inferred using bisectors. Simultaneous inversions of the \CaII\ 8542 \AA\ line and the \CaIIH\ core were performed using the NICOLE code. The nature of the oscillations and insights into the resonant oscillatory pattern were determined by analyzing the phase shift between the velocity signals and examining the temporal evolution.}
   {Propagating waves in the low chromosphere are more common in regions with frequent umbral flashes, where the transition region is shifted upward, making resonant cavity signatures less noticeable. In contrast, areas with fewer umbral flashes show velocity fluctuations that align with standing oscillations. Evidence suggests dynamic changes in the location of velocity resonant nodes due to variations in transition region height. Downflowing profiles appear at the onset of some umbral flashes, but upflowing motion dominates during most of the flash. These downflowing flashes are more common in standing umbral flashes.}
   {We confirm the existence of a chromospheric resonant cavity above sunspot umbrae produced by wave reflections at the transition region. The oscillatory pattern depends on the transition region height, which exhibits spatial and temporal variations due to the impact of the waves.}


   \maketitle
%
%-------------------------------------------------------------------

\section{Introduction}\label{sec:intro}

Umbral flashes are sudden brightenings that appear in the core of some chromospheric lines in sunspot umbrae. Since their first detection in the late 60s \citep{Beckers+Tallant1969, Wittmann1969} they have been interpreted as the result of waves propagating from the photosphere to upper chromospheric layers. Their amplitude increases due to the drop of the density \citep[e.g.,][]{Centeno+etal2006}, leading to shocks that heat the umbral chromosphere and produce a reversal in the line core, that appear in emission during the umbral flash event \citep{Havnes1970}.

Earlier studies of umbral flashes based on the analysis of spectropolarimetric observations revealed velocity fluctuations in agreement with the scenario of upward wave propagation. Using Non-Local Thermodynamic Equilibrirum (NLTE) inversions, \citet{SocasNavarro+etal2001} modeled umbral flashes as two atmospheric models inside the resolution element, including one hot upflowing component. Subsequent spectropolarimetric inversions from observations with higher spatial resolution also found umbral flashes taking place when the atmosphere exhibits strong upflows \citep{delaCruz-Rodriguez+etal2013}. This association of upflows with intensity (temperature) enhancements conforms to the accepted consensus of umbral flashes as a sign of upward propagating waves. Numerical modeling of sunspot oscillations, including the synthesis of the spectral line profiles, also supports the mechanism of propagating waves as the cause of upflowing umbral flashes \citep{Bard+Carlsson2010, Felipe+etal2014b, Felipe+etal2018a}.

However, recent NLTE inversions have identified umbral flash profiles better fitted by atmospheric models with strong downflows \citep{Henriques+etal2017, Bose+etal2019, Houston+etal2020}. The inversions by \citet{Henriques+etal2017} found that 60\% of the umbral flashes correspond to atmospheres with downflows stronger than 3 km s$^{-1}$. These results challenge the widely accepted view of umbral flashes. \citet{Felipe+etal2021a} proposed that downflowing umbral flashes can appear as a result of standing oscillations in the sunspot chromosphere. In this model, temperature enhancements produced by the standing wave partially coincide with the downflowing stage of the oscillation, and the core of the \CaII\ 8542 \AA\ can appear in emission when the atmosphere is downflowing. The chromospheric temperature keeps increasing while the atmosphere suddenly changes from downflowing to upflowing. This way, during most of the umbral flash event the atmosphere exhibits upflowing velocities. This model is based on the presence of a resonant cavity in sunspot chromospheres produced by waves trapped between the temperature gradients of the transition region and the photosphere \citep{Zhugzhda+Locans1981, Zhugzhda2008}. The existence of this cavity above sunspot umbrae has recently been confirmed \citep[see][]{Jess+etal2020, Felipe+etal2020, Felipe2021, Jess+etal2021}.

Here we present a thorough analysis of the intensity profiles of chromospheric lines in a sunspot umbra. We focus on the velocity inferences and velocity oscillatory phase to provide unique insights into the nature of umbral waves and the origin of umbral flashes. The organization of the paper is as follows: in Sect. \ref{sect:observations} we present the observed data, Sect. \ref{sect:analysis} describes the methods employed for the analysis of the observations, Sect. \ref{sect:results} shows the results, which are further discussed in Sect. \ref{sect:discussion}. Finally, the main conclusions are summarized in Sect. \ref{sect:conclusions}.

%--------------------------------------------------------------------
%--------------------------------------------------------------------
\section{Observations} \label{sect:observations}

The target of the observations is the sunspot from active region NOAA 13001. The data were acquired on 2022 May 3, when the sunspot was located at solar coordinates ($x=93\arcsec$, $y=-457\arcsec$), with the Swedish 1m Solar Telescope \citep[SST;][]{Scharmer+etal2003}. Simultaneous multi-line observations in the \CaII\ 8542 \AA, \CaIIH, \Ha, and \Hb\ lines were taken using CRisp Imaging SpectroPolarimeter \citep[CRISP;][]{Scharmer2006,Scharmer+etal2008} and CHROMospheric Imaging Spectrometer \citep[][]{Scharmer2017} instruments. Here we focus on the study of a temporal series of spectroscopic data acquired between 09:36 and 10:16 UT. The analysis is complemented by 10 full spectropolarimetric maps of the \CaII\ 8542 \AA\ line acquired immediately before the spectroscopic series, between 09:29 and 09:35 UT.

CRISP instrument was employed to scan the \CaII\ 8542 \AA\ and \Ha\ lines. The spatial sampling is 0.\arcsec0592. The diffraction limit of the SST (computed as $\lambda/D$, where $\lambda$ is the wavelength of the observed line and $D = 0.97$ m is the diameter of the telescope) is 0.\arcsec18 for \CaII\ 8542 \AA\ and 0.\arcsec14 for \Ha. The \CaII\ 8542 \AA\ was sampled with a total of 25 wavelength points. Around the core of the line (between $-390$ and $+390$ m\AA) the step was $\delta\lambda=65$ m\AA, and a coarser resolution of 2$\delta\lambda$ was used in the wings up to $\pm1040$ m\AA. Two additional points were scanned at $\pm1755$ m\AA. The \Ha\ line was sampled at 21 wavelength positions, with steps of 100 m\AA\ between -800 and +800 m\AA\ for the line center and a two times coarser resolution between $\pm800$ and $\pm1200$ m\AA. The temporal cadence of the CRISP spectroscopic series is 12.30 s.

The \CaIIH\ and \Hb\ data were acquired with CHROMIS. It has a spatial sampling of $0.\arcsec038$ and a diffraction limit of 0.\arcsec08 and 0.\arcsec10 at the wavelengths of the \CaIIH\ and \Hb\ lines, respectively. The \CaIIH\ line was sampled from -455 to +455 m\AA\ in
steps of 65 m\AA, between $\pm455$ and $\pm845$ m\AA\ in steps of 130 m\AA, and between $\pm845$ and $\pm1235$ m\AA\ in steps of 195 m\AA. An additional scan at a wavelength of 4000 \AA\ was also acquired for a total of 26 points. The \Hb\ line was scanned with the same wavelength sampling used for \Ha. The temporal cadence of the CHROMIS spectroscopic series is 12.37 s.

The data was processed using SSTRED pipeline \citep{Lofdahl+etal2021}, which employs Multi-Object Multi-Frame Blind Deconvolution \citep[MOMFBD;][]{vanNoort+etal2005,Lofdahl2002}. It implements the methods to produce science-ready data from CRISP and CHROMIS raw observations. The pipeline for the former instrument was previously described in \citet{delaCruz-Rodriguez+etal2015}.

%\begin{figure}[ht]  %%% Figure 1
%\centering
%\includegraphics[width=0.40\textwidth]{HMI.eps}
%\caption{HMI continuum intensity map of the sunspot NOAA 12662 on 2017 June 18 09:15 UT. The approximate position of the GRIS slit is shown by the black line in the HMI continuum intensity map.}\label{icplots}
%\end{figure}

%--------------------------------------------------------------------
\section{Data analysis}
\label{sect:analysis}
\subsection{Identification of umbral flash profiles}
\label{sect:identification_UFs}
In this paper, we focus on the study of umbral fluctuations. The umbral region was selected by applying an intensity threshold in the \CaII\ 8542 \AA\ maps at $\Delta\lambda=1755$ m\AA. Within the umbra, we labeled as umbral flashes those profiles where the maximum intensity between $\Delta\lambda=-325$ m\AA\ and $\Delta\lambda=0$ m\AA\ is higher than the intensity at $\Delta\lambda=650$ m\AA. Red, orange, and yellow contours in the bottom panels of Fig. \ref{fig:polarimetry_inv} illustrate the location of the umbral flashes during the temporal period analyzed in this work (last 13 min of the spectroscopic temporal series).

\subsection{Spectropolarimetric inversions}
\label{sect:polarimetry_inversion}
Full spectropolarimetric inversions of a \CaII\ 8542 \AA\ map were performed using the NLTE inversion code NICOLE \citep{SocasNavarro+etal2015}. In the presence of magnetic fields, the code calculates the polarization resulting from Zeeman splitting. The Ca II atom model employed encompasses five bound levels and a continuum \citep{delaCruz-Rodriguez+etal2012}. Collision-induced line broadening was evaluated using the approach described by \citep{Anstee+O'Mara1995}.

We constructed a sunspot map free of umbral flash profiles as follows. From the series of 10 spectropolarimetric maps, we chose the one acquired under better seeing conditions (as indicated by the Fried parameter $r_{\rm 0}$) and replaced the profiles where the core of the \CaII\ 8542 \AA\ line is in emission by profiles from the same locations scanned 69 s later. This approach provides us with a map of the quiescent umbra, where no locations are impacted by the striking changes produced by umbral flashes.  

\begin{figure*}[ht]  %%% Figure 1
\centering
\includegraphics[width=0.9\textwidth]{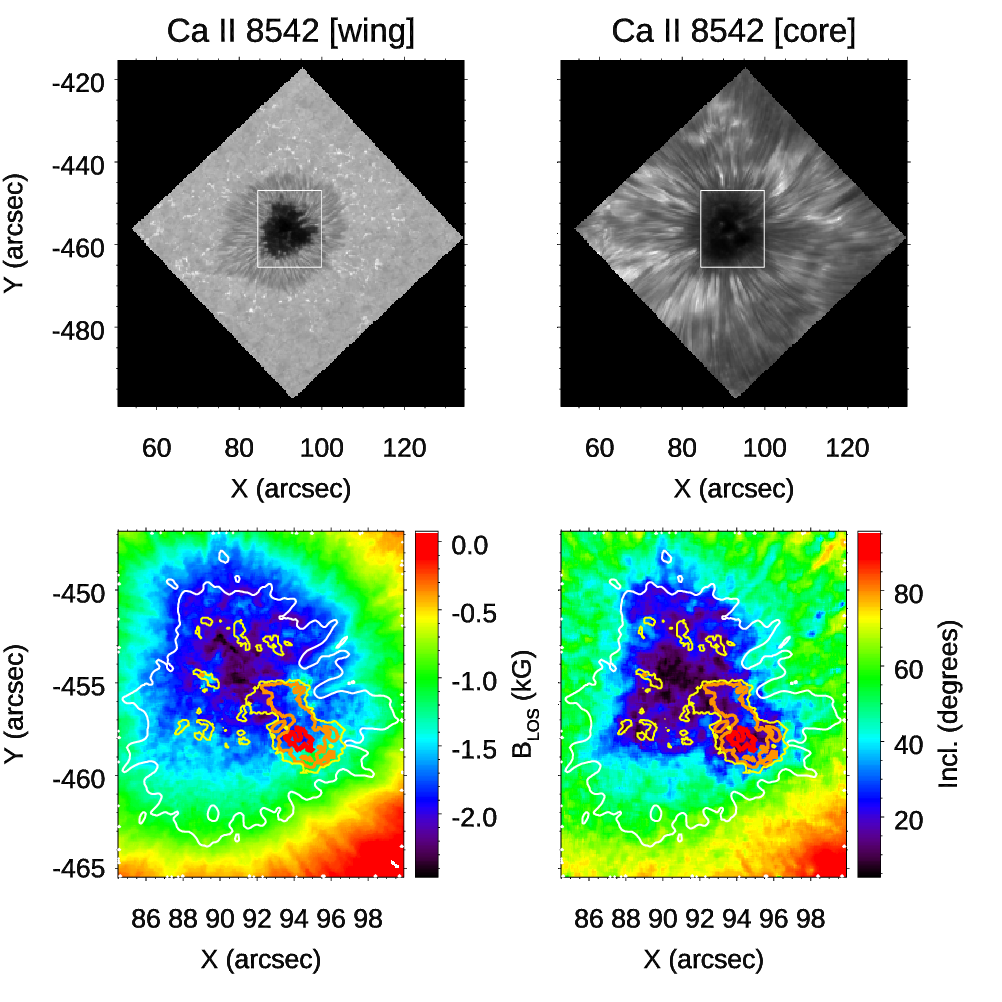}
\caption{Field of view of the observations and magnetic field inferred from the inversions of the polarimetric map. Top panels: Intensity in the wing (left panel) and core (right panel) of the \CaII\ 8542 \AA\ line. The white square delimits the region where the spectropolarimetric inversions were performed. Bottom panels: line-of-sight magnetic field (left panel) and magnetic field inclination (right panel). White contours mark the boundary between the umbra and penumbra. Red, orange, and yellow contours delimit the regions where 14, 7, and 3 umbral flash profiles, respectively, have been detected during the 13 min of the temporal series that have been analyzed. }\label{fig:polarimetry_inv}
\end{figure*}

The inversions of the polarimetric map were restricted to the sunspot umbra and surroundings. They were performed using three cycles. The last cycle includes 10 nodes for temperature, 6 for velocity, 2 for microturbulence and longitudinal magnetic field, and 1 node for the transversal components of the magnetic field. Figure \ref{fig:polarimetry_inv} illustrates the whole field of view of the observations in \CaII\ 8542 \AA\ and the inversion results for the chromospheric magnetic field along the line of sight (LOS) and inclination. A median filter, replacing the value of each element by the median found in a $5\times5$ window around the element, was applied to both inverted maps. The discrepancy between the umbra intensity contours and the LOS magnetic field strength is due to the position of the sunspot on the solar disk. Near the north umbral boundary, the inclined magnetic field lines (concerning the local reference frame) align more with the LOS and, thus, a stronger LOS magnetic field is measured there.

\subsection{Multi-line spectroscopic inversions and degeneracy of the solutions}
\label{sect:inversions}

Simultaneous inversions of the \CaII\ 8542 \AA\ and \CaIIH\ lines were performed using NICOLE. The NICOLE code operates under the assumption of full angle and frequency redistribution, along with a plane-parallel atmosphere for every pixel. This is a suitable approximation for the \CaII\ 8542 \AA\ since it is weakly affected by partial redistribution effects \citep{Uitenbroek1989}. In contrast, the \CaIIH\ and \CaIIK\ lines are formed at higher chromospheric layers, where partial redistribution is more important \citep{Vardavas+Cram1974, Uitenbroek1989, Solanki+etal1991, Bjorgen+etal2018}. These effects are especially fundamental in the wings of the lines. Previous works have independently modeled the core and the wings of some of these lines, including 3D radiative transfer and complete redistribution in the core and 1.5D radiative transfer and partial redistribution in the wings \citep{Pereira+etal2013}. Here, we have restricted the multi-line inversions to the core of the \CaIIH\ line, in addition to the full \CaII\ 8542 \AA\ profile, where the impact of partial redistribution effects is weaker. Only the wavelengths of the \CaIIH\ line between $\Delta\lambda$=-130 m\AA\ and $\Delta\lambda$=65 m\AA\ (4 spectral points) were used for the inversions. A weight two times higher than that of the \CaII\ 8542 \AA\ was assigned to those points.

\begin{figure}[ht]  %%% Figure 1
\centering
\includegraphics[width=0.48\textwidth]{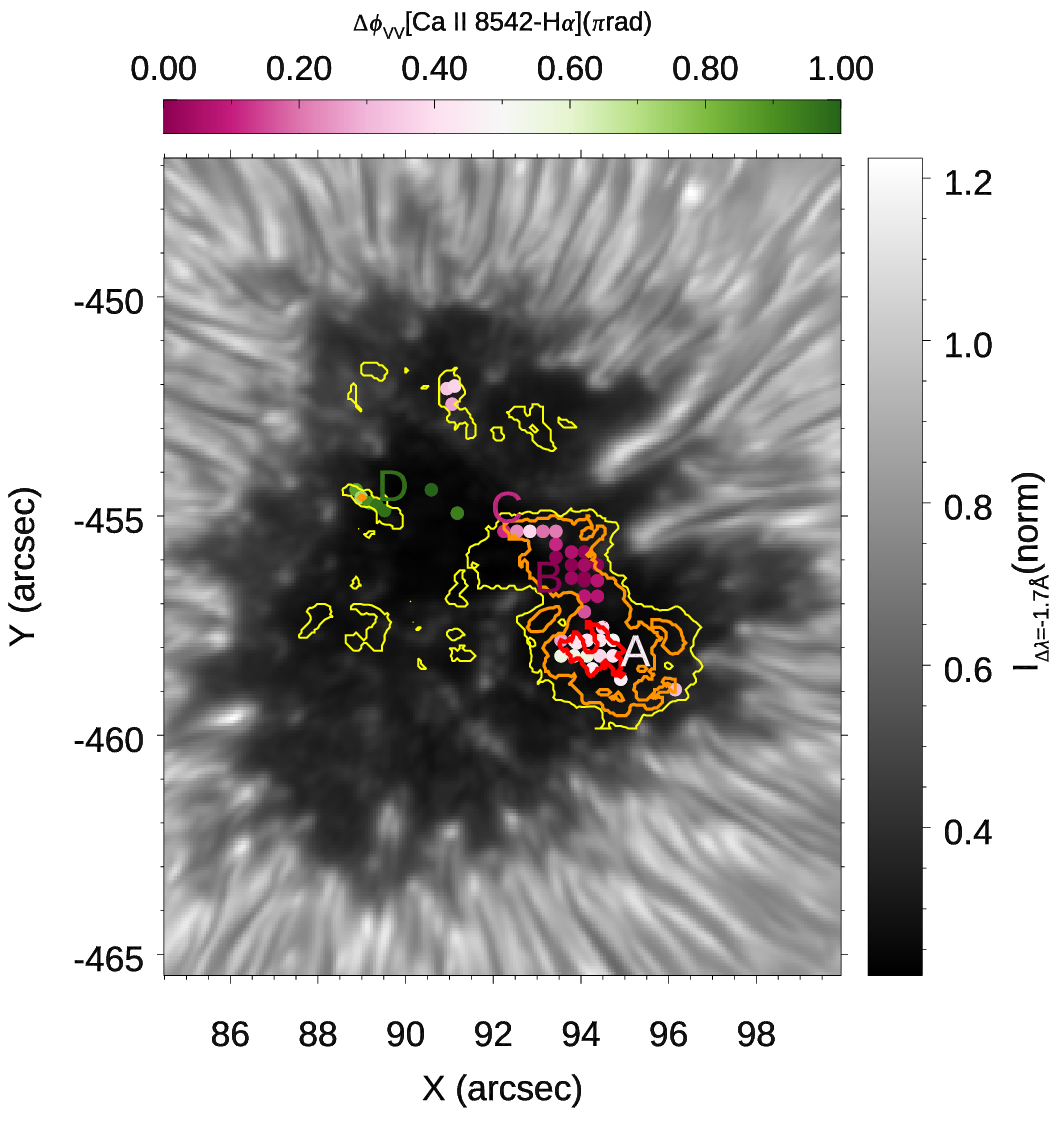}
\caption{V-V phase shift between the \CaII\ and \Ha chromospheric velocities at the analyzed locations. The background illustrates an intensity map of the umbra and surroundings at the continuum around the \CaII\ 8542 \AA\ line. The colored dots mark the analyzed locations, with the color corresponding to the V-V phase shift at 6.45 mHz according to the upper colour bar. The contours indicate the number of umbral flash profiles as described in Fig. \ref{fig:polarimetry_inv}. Letters A-D indicate the locations illustrated in Figs. \ref{fig:inv_propagating}, \ref{fig:inv_standing}, \ref{fig:inv_standing_anti},  \ref{fig:inv_standing_v0}.}\label{fig:map_dphase}
\end{figure}

The \CaII\ 8542 \AA\ and \CaIIH\ profiles were acquired with different instruments offering different spatial sampling and temporal cadence. For the inversions, we selected a \CaII\ 8542 \AA\ profile (from CRISP instrument, since it has a coarser spatial sampling) and searched the \CaIIH\ profile whose spatial location and time is closer to the chosen \CaII\ 8542 \AA\ spectrum. This means that the profiles of the two lines are not strictly co-spatial and co-temporal. In the following analyses, we have focused on 44 spatial locations to sample the sunspot umbra. These locations (illustrated by color circles in Fig. \ref{fig:map_dphase}) were selected at positions where the distance between CRISP and CHROMIS pixels is under 0$\arcsec$01. The maximum time shift between the profiles of both instruments is around 6 s. We focus the analysis on cases when the two lines are acquired nearly simultaneously, allowing a maximum time difference of roughly 2.5 s. Thanks to the similar temporal cadence of both instruments (12.30 and 12.37 s for CRISP and CHROMIS, respectively), the temporal coherence is reasonably maintained for successive scans. Following this criterion we have focused the analysis on the last 13 min of the spectroscopic series.

In order to minimize the uncertainty of the results of the NLTE inversions due to the degeneracy of the solutions \citep[\eg,][]{Felipe+etal2021b}, we have performed a total of 256 inversions from each analyzed profile. They differ in the initial guess atmosphere, the number of employed nodes, and the regularization parameter. Every independent inversion was carried out using a scheme of several cycles with increasing complexity in the node distribution. The number of nodes for the last cycle varies between 6 and 10 for the temperature, 2 and 6 for the velocity, and 1 and 2 for the microturbulence. In the inversions of the spectroscopic temporal series, the magnetic field was not inverted. Instead, the value of the magnetic field vector extracted from the inversion of the spectropolarimetric map (Sect. \ref{sect:polarimetry_inversion}) at the corresponding location was imposed. NICOLE inversions generally include regularization. This is a free parameter that can be set to values between 0 and 1. Values near 1 tend to favor vertically smooth profiles, whereas regularization near 0 will allow solution atmospheres with strong vertical variations. In the set of control files used for the inversions, the regularization ranges from 0.1 to 1. The process was parallelized using the \emph{parallel} command \citep{parallel}. 

\begin{figure}[ht]  %%% Figure 1
\centering
\includegraphics[width=0.48\textwidth]{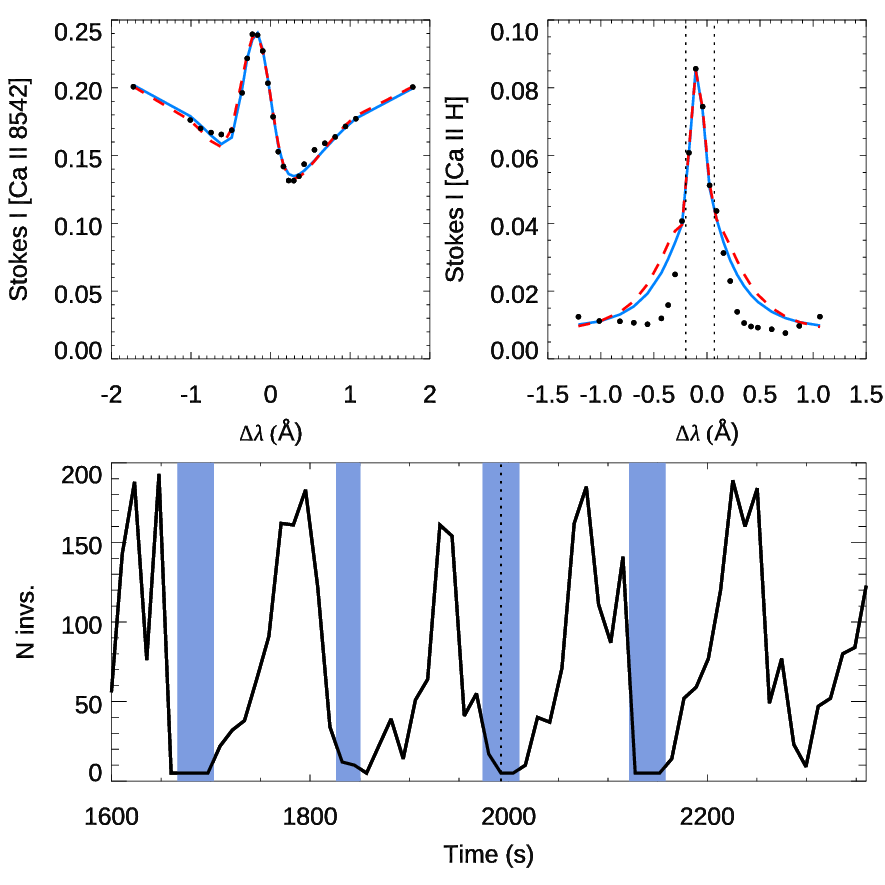}
\caption{Quality of the inversion fits and number of inversions that satisfy the threshold criterion. Top panels: observed umbral flash profile (black dots), the fit obtained from an inversion whose $\chi^2$ is near the chosen threshold (blue solid line), and the fit of the inversion with the worst $\chi^2$ (above the threshold) that was included in the analysis (red dashed line). The left panel corresponds to the \CaII\ 8542 \AA\ line and the right panel to the \CaIIH\ line. The vertical dotted lines in the right panel delimit the line core region employed in the inversions. Bottom panel: temporal evolution of the number of inversions employed for the analysis of each profile at a selected umbral location. The blue-shaded areas indicate the temporal steps when the core of the \CaII\ 8542 \AA\ is in emission. The vertical dotted line corresponds to the time step illustrated in the top panels.}\label{fig:inv_threshold}
\end{figure}

From the total of 256 independent inversions performed for each profile, we have chosen a set of best inversions. This selection is based on the $\chi^2$ value, which characterizes the difference between the synthetic profiles generated by the atmosphere obtained from the inversion and the actual observed atmosphere. Since lower $\chi^2$ value indicates a better agreement, we imposed a maximum value in $\chi^2$. In the cases where less than five inversions satisfied this threshold in $\chi^2$, we selected the 5 inversions with lower $\chi^2$ instead. The value of $\chi^2$ computed by NICOLE depends on the weights of the wavelengths and the regularization parameter and, thus, indicating the value of that number does not provide meaningful information to the reader. To illustrate the quality of the inversion fits, we have plotted an example umbral flash profile (top panels from Fig. \ref{fig:inv_threshold}). This example corresponds to a situation where only one of the inversions is below the threshold, so four out of the five selected inversions have a $\chi^2$ above that value (bottom panel from Fig. \ref{fig:inv_threshold}). The blue lines in the top panels show the synthetic profiles obtained from the inversion whose $\chi^2$ value is just below the chosen threshold, whereas the red lines indicate the profiles for the highest $\chi^2$ inversion from the five selected cases. The comparison shows that, even when only a few inversions meet our threshold criterion, the fits are still sufficiently good. In general, during umbral flash events, the inversions are more challenging and we restrict the analysis to a set of 5 to 20 inversions. For quiescent profiles, many inversions (up to $\sim$200) satisfy the threshold criterion.

\begin{figure}[ht]  %%% Figure 1
\centering
\includegraphics[width=0.48\textwidth,trim=0cm 0cm 0cm 0cm]{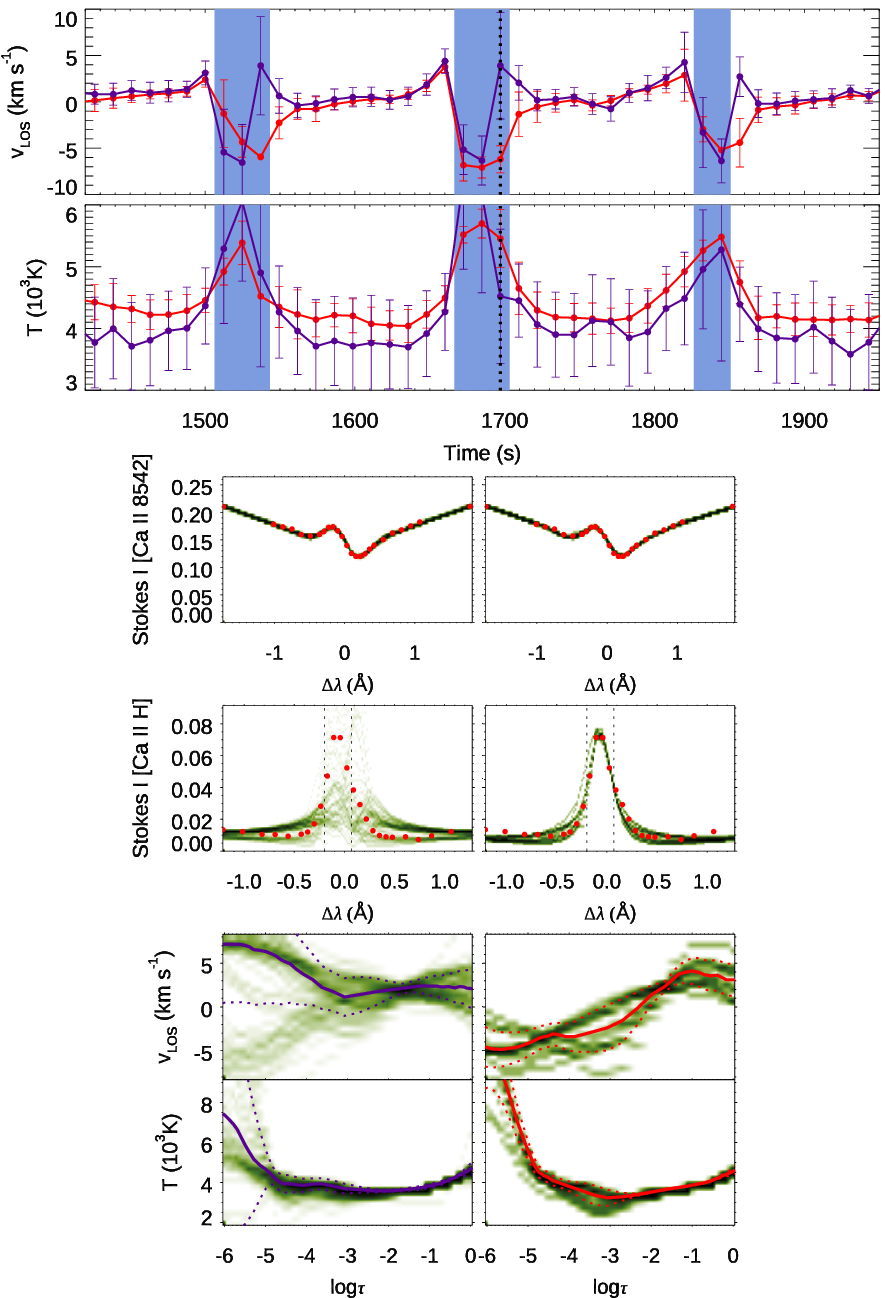}
\caption{Comparison of single-line inversions of the \CaII\ 8542 \AA\ and multi-line inversions of that line together with the core of the \CaIIH\ at a randomly chosen umbral location. Top panel: temporal evolution of the chromospheric velocity inferred from multi-line (red line) and single-line (violet) inversions. The velocity value is given by the median of the set of best inversions from the 256 independent inversions performed for each spectral profile. The error indicates the standard deviation of that set. Second panel: same as the top panel but for the temperature. In the top two panels, the blue-shaded areas indicate the times when the \CaII\ 8542 \AA\ profile exhibits an umbral flash. The vertical dotted line at t=1697 s marks the time illustrated in the bottom panels. Four bottom rows: inversion results for the single-line (left column) and multi-line (right column) inversions. The top two rows illustrate the observed line profiles (red dots) and the fit of the inversions. The bottom two rows correspond to the stratification of the velocity and temperature inferred from the inversions. In all these panels, the green color scale indicates how common is a solution in the set of best inversions (a darker color corresponds to a more common solution). In the bottom two rows, the solid lines illustrate the median solution from the set of best inversions, whereas the dotted lines indicate the standard deviation.}\label{fig:degeneracy}
\end{figure}

Figure \ref{fig:degeneracy} illustrates the degeneracy of the solutions and the necessity of including the core of the \CaIIH\ line in the inversions. It represents the results from the inversions at an umbral location, with the focus on the profiles from one temporal step with strong differences between the multi-line and single-line inversions. The same approach was followed for both cases, the only difference is that in the single-line inversions the weights of the \CaIIH\ line core were set to zero. The chromospheric velocity (averaged between $\log\tau=-4.6$ and $\log\tau=-5.3$) of the single-line inversions exhibits a good agreement with that obtained from the multi-line inversions in most of the temporal series, except around the end of the umbral flash events. At those times, the single-line inversion (violet line) shows a sudden change from negative (upflowing) to positive (downflowing) velocity. In contrast, the multi-line inversion exhibits a progressive change in the velocity from upflows towards downflows, as expected from the temporal evolution of chromospheric velocity fluctuations in sunspot umbrae \citep[\eg,][]{Centeno+etal2006}. Bottom panels illustrate the inversions for one of those cases where this discrepancy is obvious. The stratification of the velocity in the single-line inversion exhibits two independent branches, one with remarkably high negative velocities, and the other with strong positive velocities. The median velocity (violet solid line) tracks the latter branch of the solutions since most of them are gathered there. Adding the core of the \CaIIH\ line to the inversions breaks this degeneracy, discarding the solutions with positive (downflowing) velocity. It also benefits temperature estimations. During the whole temporal series, the standard deviation of the multi-line inversion temperature is significantly lower than that of the single-line inversions (second panel from Fig. \ref{fig:degeneracy}). This can also be seen in the temperature stratification (bottom panels from Fig. \ref{fig:degeneracy}), where the dispersion of the solutions in the chromosphere is much larger when only the \CaII\ 8542 \AA\ line is inverted.

\subsection{\Ha\ velocity}

The line-of-sight velocity of the \Ha\ profiles was estimated using bisectors \citep{Kulander+Jefferies1966, GonzalezManrique+etal2020}. This technique provides height-dependent velocity information by measuring the shift between the static line center and the line bisector. The measurement of the bisectors at different intensity levels, from the core to the wings of the line, gives the velocity at different atmospheric layers.  

In this work, we have derived the \Ha\ velocity for all the sunspot locations and the full temporal series using several bisectors, with intensities at 5\%, 10\%, 20\%, 30\%, 40\%, and 50\% of the line, where 0\% corresponds to the line core and 100\% to the outer wings. The velocity computed with bisectors closer to the continuum level (beyond 60\%) produced noisy velocity maps that were discarded. In the following, we will focus on the analysis of the 10\% bisectors. The other percentage line depths have also been evaluated without finding significant differences with the 10\% case.  

\subsection{V-V phase shift}

The multi-line spectroscopic inversions described in Sect. \ref{sect:inversions} have been performed for 63 time steps (around 13 min) at 44 spatial positions. The locations were selected to sample sunspot regions with different umbral flash activity. The chosen locations are marked in Fig. \ref{fig:map_dphase} by colored dots. Several locations were selected in the umbral region where flashed profiles are prevalent (more than 13 profiles during the analyzed period, red contour in Fig. \ref{fig:map_dphase}), some in the region with moderate frequency of umbral flashes (between 7 and 12 flashed profiles, orange contours) and low frequency of flashes (between 3 and 6 profiles, yellow contours). For completeness, we have also included two locations where no umbra flashes appear. 

The color of the dots indicates the V-V phase difference between the chromospheric velocity inferred from the multi-line inversion of the \CaII\ lines and the \Ha\ velocity ($\Delta\phi_{\rm VV}$[\CaII\ 8542-\Ha]) at the frequency where the power of the chromospheric velocity oscillations is maximum (6.45 mHz). The phases are directly computed from the Fourier transform of the velocity signals at the same spatial location. A phase shift of $\Delta\phi_{\rm VV}$[\CaII\ 8542-\Ha]=0 rad (dark pink) indicates that both velocity signals are in phase, whereas $\Delta\phi_{\rm VV}$[\CaII\ 8542-\Ha]=$\pi$ rad (dark green) corresponds to oscillations in opposite phase. A positive phase difference indicates a delay of the \Ha\ velocity signal concerning the \CaII\ chromospheric velocity.

\begin{figure}[ht]  %%% Figure 1
\centering
\includegraphics[width=0.48\textwidth]{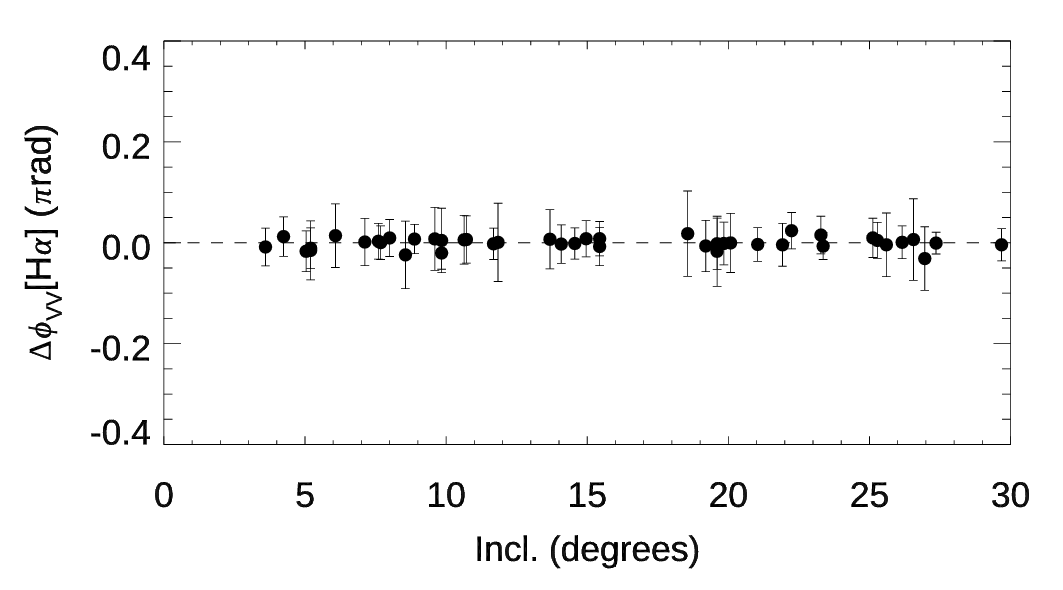}
\caption{Average phase difference between the velocity measured in H$\alpha$ at the locations analyzed in this study and the surrounding pixels within a diameter of four pixels. The results are plotted as a function of the local magnetic field inclination. The error bars indicate the standard deviation.}\label{fig:dphase_Ha}
\end{figure}

Given the height difference between the chromosphere probed by the inversions of the \CaII\ lines and the formation height of H$\alpha$ (around 265 km, see Sect. \ref{sect:propagating}) and the position of the sunspot on the solar disk, we expect a displacement of more than two pixels (from CRISP instrument) across the plane of sky between both signals. We have not taken this displacement into account when computing the V-V phase shift between the velocity signals from those two atmospheric layers. However, we can argue that this issue does not compromise the results from the phase difference analysis. We have computed the phase difference between the H$\alpha$ velocity in each selected location (44 cases) and all the surrounding positions within a diameter of four pixels. The results are illustrated in Fig. \ref{fig:dphase_Ha} as a function of the magnetic field inclination. In all cases, the average phase difference is around 0 and the standard deviation is low, indicating that H$\alpha$ velocity oscillations are coherent over spatial scales larger than the projection shift on the plane of the sky. The phase difference results between the \CaII\ 8542 \AA\ and H$\alpha$ velocities are not significantly impacted by the fine determination of the H$\alpha$ location. The situation is more complex when the propagation of slow magnetoacoustic waves (as those studied in this work) along magnetic field lines is taken into account. Depending on the orientation of the magnetic field, they could balance the projection displacement or increase it. However, the results from Fig. \ref{fig:dphase_Ha} do not depend on the inclination, proving that the measured dispersion in the H$\alpha$ velocity phase of neighboring points is not due to the propagation of the waves along the magnetic field lines.

%--------------------------------------------------------------------
\section{Standing or propagating nature of chromospheric umbral waves} \label{sect:results}

The phase shift between the chromospheric velocity inferred from the inversion of the \CaII\ lines and the \Ha\ velocity has been used to determine the standing or propagating character of the waves at each analyzed location. Standing waves oscillate in time, but the peak amplitude of the wave does not move in space. Instead, they form a stationary wave pattern, with specific points of minimum amplitude (nodes) and maximum amplitude (antinodes). Locations in the same region of the wave pattern (for example, between two nodes) oscillate in phase, whereas two points with an odd number of nodes between them oscillate in anti-phase. Propagating waves move in space. For example, in the lower solar atmosphere, the wavefronts progressively move upwards. Each point throughout the wave has a different phase, as opposed to standing oscillations, and the measurement of the phase difference between the fluctuations at two layers can be used to track and characterize this wave propagation. 

The phase shifts between the chromospheric velocity signals illustrated in Fig. \ref{fig:map_dphase} exhibit various values depending on the location. This points to the inhomogeneous nature of the chromospheric waves at the layers probed by the \CaII\ lines and \Ha. Also, the highly dynamic umbral chromosphere, with the umbral flashes as its main feature, produces shifts in the transition region and formation height of the lines. This way, at certain locations the oscillations can exhibit different behaviors. In the following subsections, we illustrate these insights by showing the results at some selected locations.

\subsection{Propagating waves in regions with high frequency of umbral flashes}
\label{sect:propagating}

\begin{figure}[ht]  %%% Figure 1
\centering
\includegraphics[width=0.48\textwidth]{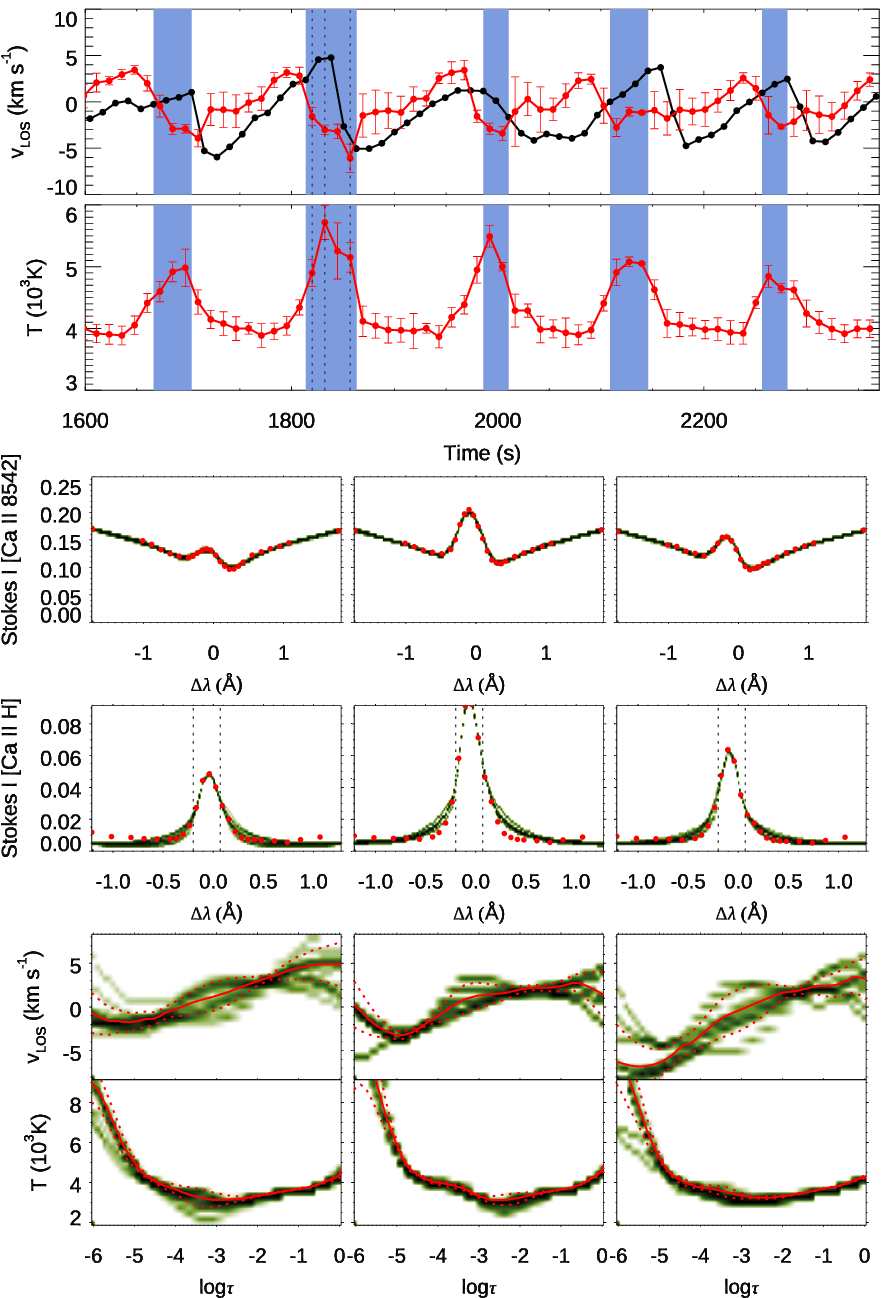}
\caption{Temporal evolution of the chromospheric fluctuations and inversion results at point A from Fig. \ref{fig:map_dphase}, where propagating waves are found in the low chromosphere. Top panel: chromospheric velocity inferred from the multi-line inversions of the \CaII\ lines (red line) and from \Ha\ (black line). The inverted velocity corresponds to the median of the set of best inversions (out of 256 independent inversions) averaged in the optical depths between $\log\tau=-4.6$ and $\log\tau=-5.3$. Error bars indicate the standard deviation of the set of best inversions. Second panel: chromospheric temperature inferred from the inversion of the \CaII\ lines. The median temperature and error bars were determined similarly to the velocity. Middle panels: intensity of the \CaII\ 8542 \AA\ (third row) and the \CaIIH\ (fourth row) lines at three time steps indicated by vertical dashed lines in the top two panels. Bottom panels: inversion results for the velocity (fifth row) and temperature (last row) for the three selected time steps. The green color scale has the same meaning as in Fig. \ref{fig:degeneracy}.}\label{fig:inv_propagating}
\end{figure}

Figure \ref{fig:inv_propagating} shows the temporal evolution of the velocity and temperature chromospheric fluctuations at the location A indicated in Fig. \ref{fig:map_dphase}. In the following, the discussion of the inverted velocity and temperature signals will refer to the averages in the optical depth range between $\log\tau=-4.6$ and $\log\tau=-5.3$. Location A corresponds to a selected location within the region where umbral flash profiles are most common (more than 13 flashed profiles in the analyzed series). In this region, the phase shift between both velocity signals is around 0.4$\pi$ (represented by a light pink color in Fig. \ref{fig:map_dphase}). This phase difference corresponds to a delay of around 31 s in the \Ha\ velocity signal concerning the \CaII\ chromospheric velocity, as seen in the top panel of Fig. \ref{fig:inv_propagating}. Since the wavefronts require more time to reach the higher formation heights of the \Ha\ line, this result is consistent with the presence of upward propagating waves between the layers probed by the \CaII\ lines and \Ha. Assuming that waves are free to propagate vertically at the chromospheric sound speed \citep[around 8.5 km s$^{-1}$][]{Maltby+etal1986} at all locations between these two layers, this delay corresponds to a height difference of roughly 265 km between the response height of both velocity signals. 

\begin{figure}[ht]  %%% Figure 1
\centering
\includegraphics[width=0.48\textwidth,trim=0cm 0cm 0cm 0cm]{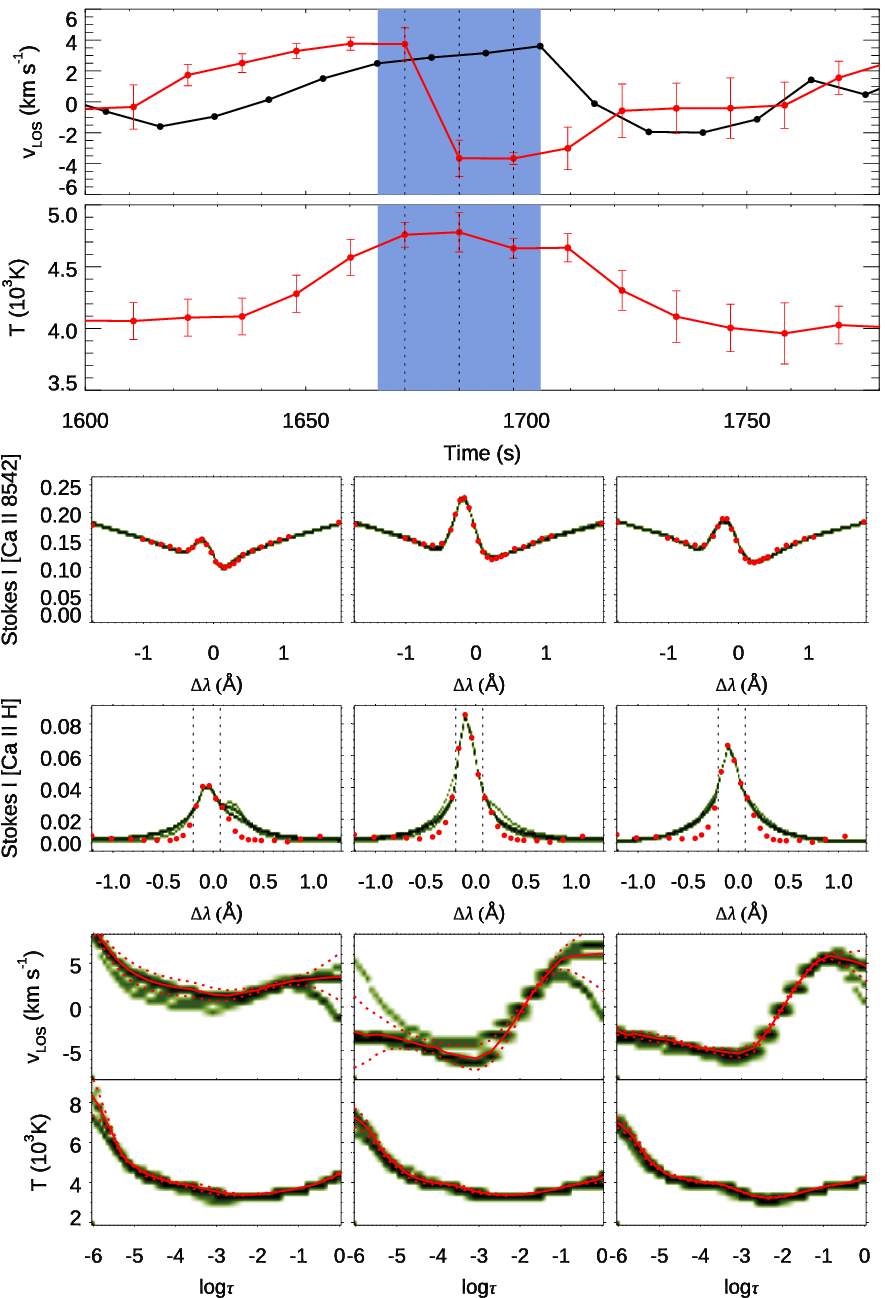}
\caption{Same as Fig. \ref{fig:inv_propagating} but for a case where a downflowing umbral flash profile is found.}\label{fig:downflowing}
\end{figure}

The region with a high number of umbral flash profiles (locations where we identify more than 13 flashed profiles, red contour in Fig. \ref{fig:map_dphase}) was sampled at 14 locations. During the 13 min of the temporal series analyzed, 5 independent umbral flash events took place at every location, as seen in the top panels of Fig. \ref{fig:inv_propagating}. The duration of umbral flash events is variable. The longest cases exhibit flashed profiles in \CaII\ 8542 \AA\ during 5 consecutive time steps (61.5 s). Almost all umbral flashes are associated with negative velocities (upflows), except for a few exceptions. Figure \ref{fig:downflowing} illustrates one of these exceptions. The \CaII\ 8542 \AA\ core is in emission for three consecutive time steps (36.9 s). At the first of them, the chromospheric velocity inferred from the inversion of the \CaII\ lines exhibits a strong downflow. As the evolution of the umbral flash event continues, the \CaII\ velocity sharply changes from positive to negative (downflow to upflow) as the core emission is enhanced.

\subsection{Umbral flashes produced by standing waves: velocity fluctuations in phase}

The phase shifts illustrated in Fig. \ref{fig:map_dphase} show a clear difference between the locations in the region with high number of umbral flashes (red contour) and the surroundings. The analyzed positions located slightly further north exhibit a phase shift around 0, pointing to \CaII\ and \Ha\ velocity fluctuations in phase. Taking into account the height difference between the probed velocity signals (as seen in the previous section), this indicates that oscillations in those locations are produced by mostly standing waves.

\begin{figure}[ht]  %%% Figure 1
\centering
\includegraphics[width=0.48\textwidth]{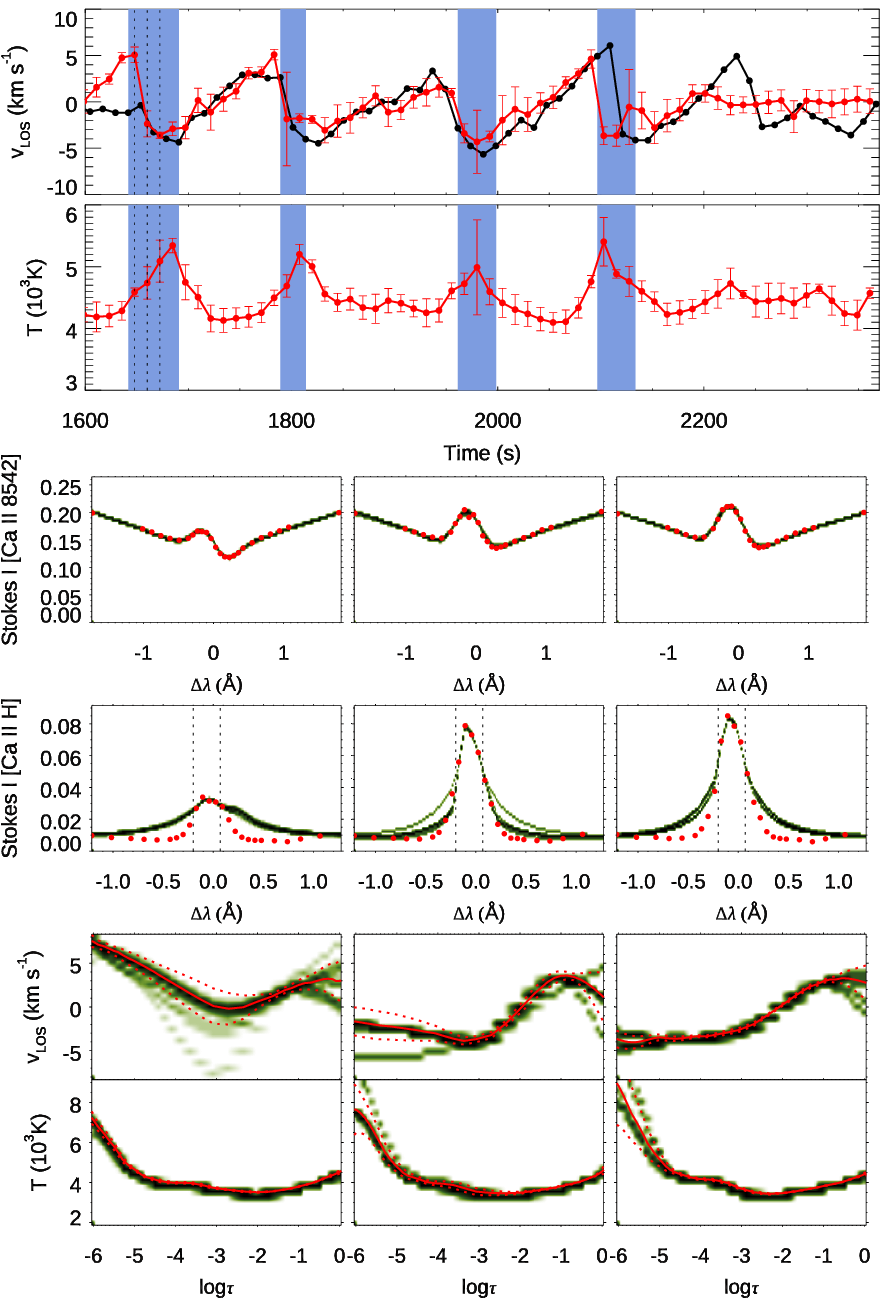}
\caption{Same as Fig. \ref{fig:inv_propagating} but for point B from Fig. \ref{fig:map_dphase}, where standing waves are found in the low chromosphere.}\label{fig:inv_standing}
\end{figure}

Figure \ref{fig:inv_standing} shows the results at the position marked as B in Fig. \ref{fig:map_dphase}, one of those locations where we detect standing waves. During most of the temporal series, especially between $t\sim1660$ s and $t\sim2100$ s, both velocity signals fluctuate in phase. The figure shows the spectral profiles and inversions during the first umbral flash event in the analyzed temporal series. In this case, the umbral flash begins when the chromospheric layer probed by the \CaII\ lines is downflowing, and continues with a sudden change to upflows, similarly to the propagating case illustrated in Fig. \ref{fig:downflowing}. Although the majority of umbral flash events are associated with upflowing chromospheres at all time steps with a \CaII\ 8542 \AA\ emission core, in the case of standing oscillations the evolution from downflowing to upflowing during the flash is more common. Out of the 57 independent umbral flash events produced by standing oscillations that we have sampled, 13 exhibit a downflowing chromosphere at the initial flashed profile. In the case of propagating waves explored in the previous subsection, only 3 out of 64 events show this behavior.

\subsection{Umbral flashes produced by standing waves: velocity fluctuations in anti-phase}

In the region around [$x=89\arcsec$, $y=-455\arcsec$] we find a moderate amount of umbral flashes profiles where the phase difference between the two chromospheric velocity signals is around $\pi$, that is, the chromospheric velocity fluctuations in the \CaII\ lines and \Ha\ are approximately in opposite phase. We also identify velocity fluctuations in anti-phase at the two sampled locations at regions free of umbral flash profiles, corresponding to the darker part of the umbra where the magnetic field is stronger.

\begin{figure}[ht]  %%% Figure 1
\centering
\includegraphics[width=0.48\textwidth]{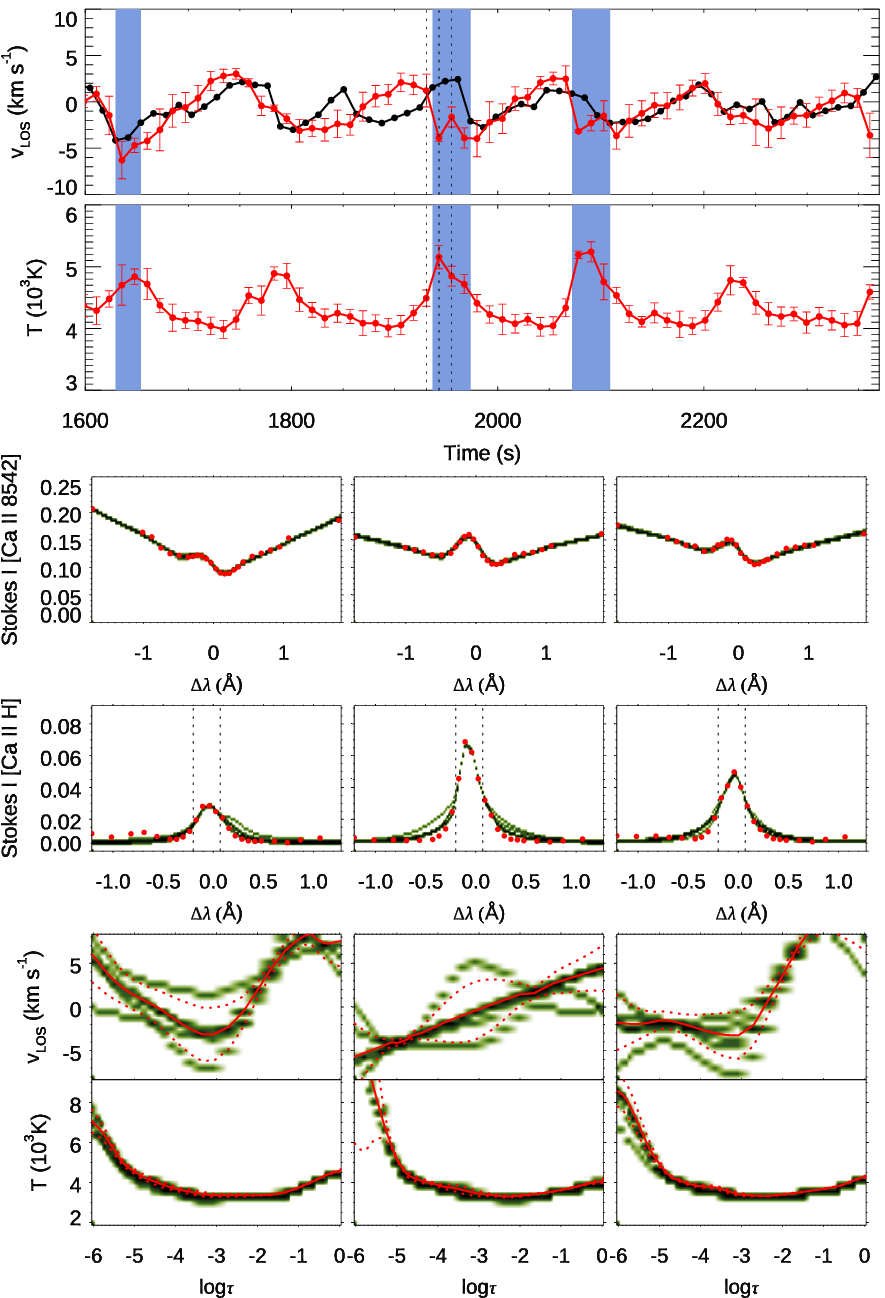}
\caption{Same as Fig. \ref{fig:inv_propagating} but for point C from Fig. \ref{fig:map_dphase}, where standing waves are found in the low chromosphere.}\label{fig:inv_standing_anti}
\end{figure}

Figure \ref{fig:inv_standing_anti} illustrates the results at an interesting location in that region (location C in Fig. \ref{fig:map_dphase}). Even though the velocity inferred from the inversions exhibits some dispersion, the solutions converge at the chromospheric height used for the velocity signal illustrated in the top panel ($\log\tau$ between -4.6 and -5.3). The velocity fluctuations are in phase at some times, for example after $t=2100$ s and at the initial times (between $t=1600$ and $t=1750$ s). This is consistent with the results presented in the previous subsection, where velocity fluctuations in phase at different low chromospheric layers indicate that oscillations are mostly produced by standing waves.  

In contrast, between $t\sim1800$ and $t\sim2000$ s (approximately for one and a half period) both velocity signals fluctuate with opposite phases. This result is also consistent with the presence of standing oscillations. We expect to find this behavior when the heights of the chromospheric velocities probed with the inversions of the \CaII\ lines and with \Ha\ are at different sides of a velocity resonant node. The height of the inferred velocities concerning the resonant node will result from an interplay between changes in the formation height of the lines and dynamic variations in the height of the transition region. We have identified different behaviors in the V-V phase shifts between the \Ha\ y \CaII\ velocities (propagating waves, standing waves in phase, standing waves in opposite phase), even though the chromospheric inferences (velocity and temperature evolutions) are similar. This way, we consider that the main driver of the changes in the oscillatory pattern is the height of the transition region.

Our results point out that the transition region height is not only inhomogeneous across the umbra, but also highly dynamic. A fixed spatial location can exhibit remarkable changes in the height of the reflecting layer producing the resonant cavity and, thus, the height of the resonant nodes.

\subsection{Umbral flashes produced by standing waves: \CaII\ 8542 \AA\ core reversal with vanishing velocity}

\begin{figure}[ht]  %%% Figure 1
\centering
\includegraphics[width=0.48\textwidth]{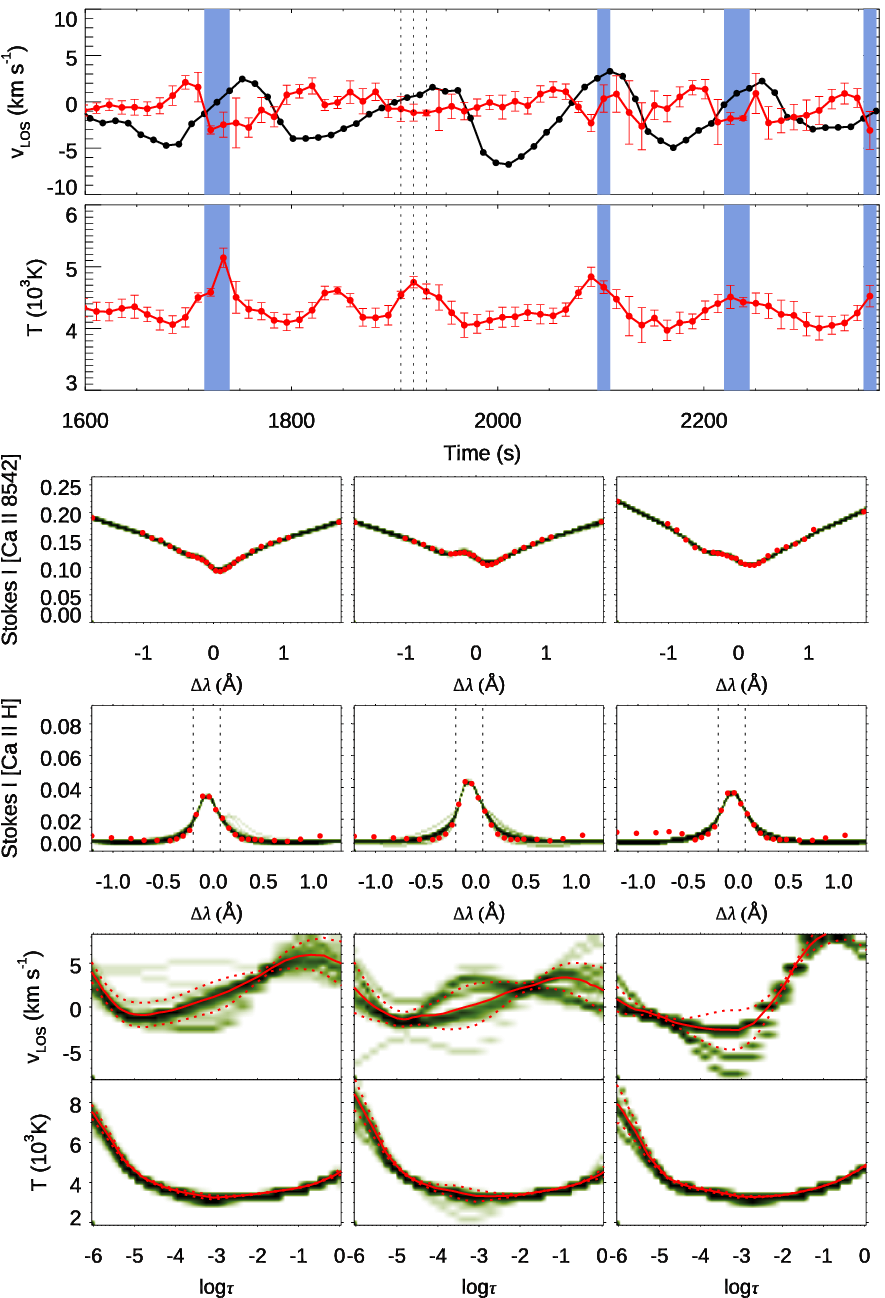}
\caption{Same as Fig. \ref{fig:inv_propagating} but for point D from Fig. \ref{fig:map_dphase}, where standing waves are found in the low chromosphere.}\label{fig:inv_standing_v0}
\end{figure}

Figure \ref{fig:inv_standing_v0} presents another example of a location (location D in Fig. \ref{fig:map_dphase}) where the velocity signals are predominantly in anti-phase. This observation is consistent with standing oscillations, indicating that the probed heights are on opposite sides of a velocity node. Interestingly, between $t=1900$ and $t=2030$ s the \Ha\ velocity exhibits remarkable velocity fluctuations with peak-to-peak amplitude of approximately 9 km s$^{-1}$. In contrast, the \CaII\ velocity signal is absent, showing a velocity around 0 for nearly an entire oscillatory period. However, during this time, the temperature fluctuations display a pronounced peak.

The oscillatory behavior observed during this temporal period also supports the presence of standing oscillations. At the height of the velocity nodes, the velocity fluctuations have zero amplitude. In this example, for some time the chromospheric height probed by the inversions of the \CaII\ lines corresponds to the height of a velocity node. The \Ha\ line, formed a few hundred kilometers above and away from the velocity node, does exhibit significant velocity fluctuations. We emphasize that the height difference between the response of both velocity signals is not sufficient to explain the large difference in the amplitude of the observed velocity if it were produced by the drop of the density between the two layers. The presence of a velocity node at the formation height of the \CaII\ lines is the only scenario that can account for these observations.

We also point out that in the chromospheric resonant cavity, the height of the velocity resonant nodes differs from the temperature resonant nodes \citep[see, e.g., Fig. 3 from][]{Fleck+Deubner1989}. Even if we assume that the velocity and temperature signals obtained from the inversion of the \CaII\ lines are probing similar atmospheric layers, this region does not correspond to a temperature resonant node. As a result, our observations exhibit a temperature peak (at the vertical dashed lines from the second panel of Fig. \ref{fig:inv_standing_v0}, illustrated in the middle and bottom panels of the same figure) that is not associated with a velocity counterpart. Interestingly, the \CaII\ 8542 \AA\ profiles during these times (third row from Fig. \ref{fig:inv_standing_v0}) show a reversal in the core of the line. Although it is too weak to be labeled as an umbral flash according to our criteria (defined in Sect. \ref{sect:identification_UFs}), these observations show that core reversals can even be found with velocities around 0, and that the temperature enhancement produced by the waves is sufficient to produce the reversal, without the presence of shocks.

%--------------------------------------------------------------------

\section{Discussion}\label{sect:discussion}

In this paper, we have studied waves in the umbral chromosphere using observations in \Ha, \CaII\ 8542 \AA, and \CaIIH. Our analysis focuses in the nature of the oscillations associated with umbral flashes. With this aim, we have performed simultaneous NLTE inversions of the two \CaII\ lines, and we have determined the velocity fluctuations from \Ha.

\subsection{Umbral flashes in the chromospheric resonant cavity}

The V-V phase difference between the velocity signal inferred from the inversions of the \CaII\ lines and that obtained from \Ha\ have been employed to evaluate the propagating or standing nature of the waves. In the umbral region where umbral flashes are more common, we find evidence of wave propagation between the chromospheric layer probed by the \CaII\ lines and the response height of \Ha. This contrasts with other umbral regions, where we find numerous evidences of the presence of mostly standing oscillations at those same chromospheric heights. These insights include the detection of \CaII\ and \Ha\ velocities fluctuating in phase, consistent with a standing wave where both velocity signals originated at the same region of the resonant pattern (with no velocity nodes between them). Also, we have analyzed some locations where the velocity signals are mostly in anti-phase, pointing to the presence of a velocity node in between their response heights, and even an instance where the \CaII\ chromospheric velocity vanishes due to the coincidence of its response height with the location of a velocity resonant node.

These different behaviors regarding the propagating or standing nature of the waves at the probed layers and the V-V phase relations are mainly due to the changes in the transition region height. Recent works have shown evidence of the existence of a resonant cavity above sunspot umbrae \citep{Jess+etal2020, Felipe+etal2020}, produced by the strong temperature gradients at the photosphere and transition region. Numerical modeling from \citet{Felipe+etal2020} showed that propagating and standing waves co-exist in the umbral chromosphere. Waves propagate from the deep photosphere to the upper layers. The partial reflection of waves at the sharp temperature gradients from the transition region produces a standing wave pattern from the interference of the upward propagating waves with the downward reflected waves. Since some of the waves reaching the transition region are leaked into the corona, there is a net upward wave flux rather than a perfect standing wave pattern. The oscillations are mostly standing in the high chromosphere, closer to the reflecting transition region, and exhibit propagating features as we consider lower chromospheric layers.

The atmospheric height where waves change from mostly propagating to mostly standing depends mainly on the height of the transition region. Our observations reveal that at the lower chromosphere where the response of the \CaII\ lines is maximum (around $\log\tau=-5$), oscillations are mostly standing. Only in the region where there are many umbral flashes, that is, where there is stronger wave driving coming from the photosphere, we find indications of propagating waves between the chromospheric heights probed by the \CaII\ lines and \Ha. This is consistent with a higher transition region at locations with stronger wave flux, indicating that waves play a fundamental role in determining the transition region's height. In our observations, a high number of umbral flashes are at the west of a region with many umbral dots (see Fig. \ref{fig:polarimetry_inv}). The umbral flash profiles appear to propagate horizontally, moving away from the umbral dots region towards the outer parts of the umbra. We interpret this as a visual pattern of waves propagating from the photosphere to the lower chromosphere along the field lines, whose arrival time is delayed where the magnetic field lines are more inclined, similar to the common understanding of running penumbral waves. Our results indicate that the origin of the additional wave flux that pushes up the transition region is the magnetoconvection taking place at the umbral dots, which generates 3-minute oscillations \citep{Chae+etal2017} 

Our observations also show that the transition region height not only depends on the umbral location but also changes dynamically at a fixed position. This is supported by the finding of locations where the V-V phase shift switches between in-phase and anti-phase oscillations, or the temporal vanishing of chromospheric velocity fluctuations in \CaII\ when their response coincides with the height of a velocity resonant node. These insights point to the displacement of the resonant pattern, driven by variations in the height of the transition region. The confirmation of a chromospheric resonant cavity above sunspot umbrae has led to proposals for utilizing Fourier analysis to conduct seismological studies of the umbral chromosphere \citep{Jess+etal2020, Felipe+etal2020}. The interpretation of these seismic analyses is challenged by the temporal scales of the changes in the transition region height. Figure \ref{fig:inv_standing_anti} shows that this temporal scale can be of the order of minutes, whereas the acquisition of an observation with good enough frequency resolution in the Fourier space requires significantly longer time series.

Numerous observational and theoretical studies have interpreted umbral flashes as the result of upward shock propagation \citep[\eg, ][to name a few]{SocasNavarro+etal2001, RouppevanderVoort+etal2003, Bard+Carlsson2010, delaCruz-Rodriguez+etal2013, Felipe+etal2014b, Houston+etal2018}. Recently, \citet{French+etal2023} employed data from the NSF Daniel K. Inouye Solar Telescope \citep[DKIST][]{Rimmele+etal2020} to determine the properties of those shocks during an umbral flash event, including their Match number and propagation speed. The shock interpretation is apparently conflicting with the standing oscillations reported in this work. However, we argue that both scenarios are consistent and co-exist. At the low chromospheric layer probed by our inversions, we identify propagating shock waves, as indicated by the delay of the shock front measured at higher chromospheric layers (H$\alpha$), in those regions where umbral flashes are more common (Fig. \ref{fig:inv_propagating}). That is, many of the observed umbral flashes are generated by wave fronts that are still propagating at the formation height of the \CaII\ 8542 \AA, which is the spectral line employed by most studies of this phenomenon. These waves appear as propagating shocks in the low chromosphere but turn into standing oscillations at higher chromospheric layers, closer to the transition region. In this work, we have paid attention to another kind of umbral flash. They appear in regions with lower wave power and, thus, lower transition region height. As a result, the standing nature of the oscillations manifests at the low chromosphere probed by the \CaII\ 8542 \AA\ spectral line.

\subsection{Downflowing umbral flashes}

This paper also sheds light on the recent controversy regarding the existence of umbral flashes that are better described by downflowing atmospheres \citep{Henriques+etal2017, Bose+etal2019}. Our scanning and inversion strategy allows us to study the evolution of umbral flashes with a high temporal cadence and with higher precision than previous works. We have found a significant number of umbral flash events where the first flashed profile in \CaII\ 8542 \AA\ is associated with a downflowing chromosphere (22.8\% of the standing umbral flashes, 4.7\% of the propagating umbral flashes). Then, the evolution of the umbral flash continues, and the chromospheric velocity suddenly changes from downflowing to upflowing while the core of the \CaII\ 8542 \AA\ line is still reversed. Capturing the initial downflowing umbral flash depends on the properties of the event (how strong is the core reversal at the initial stages), the timing of the scanning concerning the development of the umbral flash, and the identification criterion for umbral flash profiles (Sect. \ref{sect:identification_UFs}). The prevalence of downflowing umbral flashes in the case of standing oscillations agrees with the predictions from the numerical modeling by \citet{Felipe+etal2021a}.

Our analyses indicate that downflowing umbral flashes, while common, are a large minority of the total number of umbral flashes. This result is a discrepancy with the inversions from \citet{Henriques+etal2017}, who found that most umbral flash profiles (around 60\%) were better fitted with downflowing solutions. They discussed that the downflowing family of solutions could be a radiative transfer effect with no counterpart in the solar atmosphere. Our analysis of the degeneracy of the inversions supports this idea since downflowing solutions are obtained when only the \CaII\ 8542 \AA\ is inverted. The addition of the \CaIIH\ core to the inversions helps to discard those solutions, keeping only the upflowing atmospheres that also offer a better match with the expected temporal evolution of the chromospheric velocity (Fig. \ref{fig:degeneracy}). However, \citet{Chae+etal2023} has recently confirmed the presence of intensity peaks during downflowing phases of umbral oscillations from observations in \Ha. The existence of mostly downflowing umbral flashes remains an open question and possibly will depend on the sunspot properties.

\section{Concluding remarks} \label{sect:conclusions}

We have performed a detailed analysis of the chromospheric oscillations in the umbra, with special attention to the umbral flashes, in high temporal cadence spectroscopic observations. The chosen inversion strategy, while highly computationally demanding, offers a more robust insight into the temporal evolution of the chromospheric oscillations than previous analyses. 

Our study confirms the existence of a chromospheric resonant cavity above sunspot umbrae, and also provides novel results concerning the nature of the resonant pattern. We have found that at the lower chromosphere oscillations are mostly standing, except in regions with high frequency of strong waves. In those regions, waves shift the transition region to upper layers, and the effects of the resonant cavity are not noticeable at the low chromosphere, which exhibits hints of upward wave propagation. The impact of waves in the transition region is also found in locations where standing oscillations are detected at the formation height of the \CaII\ lines. We have proven the dynamic change of the transition region height from the variations in the V-V phase relations between the chromospheric height probed by the inversion of the \CaII\ lines and \Ha. At a fixed location, the V-V phase can switch between in-phase and anti-phase oscillations as a result of the displacement of the transition region and, thus, changes in the atmospheric height of the resonant nodes.

We have also clarified the existence of downflowing umbral flashes. In the early stages of an umbral flash event, the \CaII\ 8542 \AA\ can exhibit an emission core when the chromospheric velocity is still positive (downflowing). Then, the velocity suddenly changes from positive to negative and remains upflowing for the rest of the event, so most of the umbral flash profiles appear during chromospheric upflows. This behavior is more commonly found in locations where standing oscillations are found at the formation height of the \CaII\ 8542 \AA\ line. The use of multi-line inversions, combining the \CaII\ 8542 \AA\ and \CaIIH\ lines, proved critical to discriminate among the solutions obtained from the single-line inversions of the \CaII\ 8542 \AA\ line. In many cases, downflowing umbral flash solutions obtained from single-line inversions of the \CaII\ 8542 \AA\ were discarded when adding the \CaIIH\ line to the inversions.

\begin{acknowledgements}
     The authors would like to express their gratitude to Mats L{\"o}fdahl for his invaluable support in processing the SST data. Financial support from grant PID2021-127487NB-I00, funded by MCIN/AEI/ 10.13039/501100011033 and by “ERDF A way of making Europe”, and from grant CNS2023-145233 funded by MICIU/AEI/10.13039/501100011033 and by “European Union NextGenerationEU/PRTR” is gratefully acknowledged. TF acknowledges grant RYC2020-030307-I funded by MCIN/AEI/ 10.13039/501100011033 and by “ESF Investing in your future”. SJGM acknowledges grant RYC2022-037565-I funded by MCIN/AEI/10.13039/501100011033 and by "ESF Investing in your future". DM acknowledges support from the Spanish Ministry of Science and Innovation through the grant CEX2019-000920-S of the Severo Ochoa Program.

\end{acknowledgements}

\bibliographystyle{aa} % style aa.bst
\bibliography{biblio.bib} % your references Yourfile.bib
\end{document}